\documentclass[12pt,preprint]{aastex}

\def\suzaku{{\it Suzaku }} 
\def\xmm{{\it XMM-Newton }}

\def\ltsima{$\; \buildrel < \over \sim \;$}
\def\simlt{\lower.5ex\hbox{\ltsima}} 
\def\gtsima{$\; \buildrel > \over \sim \;$}
\def\simgt{\lower.5ex\hbox{\gtsima}} 

\begin{document}

\title{
Low energy cut-offs and hard X-ray spectra in high-$z$ radio-loud quasars:
the {\it Suzaku} view of RBS~315
}

\normalsize \author{F. Tavecchio\altaffilmark{1},
L. Maraschi\altaffilmark{2}, G. Ghisellini\altaffilmark{1}, 
J. Kataoka\altaffilmark{3}, L. Foschini\altaffilmark{4}, 
R.M. Sambruna\altaffilmark{5}, G. Tagliaferri\altaffilmark{1}}

\altaffiltext{1}{INAF/Osservatorio Astronomico di Brera, via Bianchi
46, 23807 Merate (LC), Italy} \altaffiltext{2}{INAF/Osservatorio
Astronomico di Brera, via Brera 28, 20121 Milano, Italy}
\altaffiltext{3}{Tokio Institute of Technology, Meguro, Tokio,
152-8551, Japan} \altaffiltext{4}{INAF/IASF-Bologna, Via Gobetti 101,
40129 Bologna, Italy} \altaffiltext{5}{NASA Goddard Space Flight
Center, Code 661, Greenbelt, MD 20771, USA}

\begin{abstract} 
We present the results from the \suzaku observation of the powerful
radio-loud quasar RBS~315 ($z=2.69$), for which a previous {\it
XMM-Newton} observation showed an extremely flat X-ray continuum up to
10 keV (photon index $\Gamma=1.26$) and indications of strong
intrinsic absorption ($N_H\sim 10^{22}$ cm$^{-2}$ assuming neutral
gas). The instrument for hard X-rays HXD/PIN allows us a detection of
the source up to 50 keV. The broad-band continuum (0.5-50 keV) can be
well modeled with a power-law with slope $\Gamma =1.5$ (definitively
softer than the continuum measured by {\it XMM-Newton}) above 1 keV
with strong deficit of soft photons. The low-energy cut-off can be
well fitted either with intrinsic absorption (with column density
$N_H\sim 10^{22}$ cm$^{-2}$ in the quasar rest frame) or with a break
in the continuum, with an extremely hard ($\Gamma =0.7$) power-law
below 1 keV. We construct the Spectral Energy Distribution of the
source, using also optical-UV measurements obtained through a
quasi-simultaneous {\it UVOT/SWIFT} observation. The shape of the SED
is similar to that of other Flat Spectrum Radio Quasars (FSRQs) with
similar power, making this source an excellent candidate for the
detection in $\gamma$-rays by {\it GLAST}. We model the SED with the
synchrotron-Inverse Compton model usually applied to FSRQs, showing
that the deficit of soft photons can be naturally interpreted as due
to an {\it intrinsic curvature} of the spectrum near the low energy
end of the IC component rather than to intrinsic absorption, although
the latter possibility cannot be ruled out. We propose that in at
least a fraction of the radio-loud QSOs at high redshift the cut-off
in the soft X-ray band can be explained in a similar way. Further
studies are required to distinguish between the two alternatives.
\end{abstract} 

\keywords{Galaxies: active --- galaxies: jets --- (galaxies:) quasars:
individual (RBS~315) --- X-rays: galaxies}

\section{Introduction}

Blazars represent the most active side of Active Galactic Nuclei due
to a priviledged aspect angle enhancing the emission from the jet with
respect to that from the accretion disk. The strongly variable
non-thermal continuum, extending from the radio to the high-energy
$\gamma $--rays, produced within the jet (with bulk Lorentz factors
$\Gamma \sim 10$) is strongly amplified in the forward direction by
relativistic beaming. The short variability timescale (down to hours)
indicates that the emission region must be located quite close to the
central engine, at distances of the order of 0.01--0.1 pc. Therefore,
studying emission from blazars we can probe the physical state of the
jet close to its base, allowing important insights on the physical
mechanism able to produce such extreme structures. The double-humped
Spectral Energy Distribution displayed by blazars (Sambruna et
al. 1996, Fossati et al. 1998) is generally described in terms of
synchrotron (contributing from the radio to the UV-X-ray range) and
Inverse Compton (extending from X-rays to $\gamma$-rays) emission from
a single population of relativistic electrons (e.g., Ghisellini et
al. 1998). The peaks of the two components show a clear trend with the
total emitted power (Fossati et al. 1998). For low power sources (
bolometric luminosity $<10^{45}$ erg/s) the synchrotron peak falls in
the UV - X-rays and the IC peak in the multi GeV-TeV range. In high
power sources (bolometric luminosity $>10^{46}$ erg/s), the
synchrotron peak is located in the submm-FIR band, while the IC
component, usually dominating the total radiative output, peaks in the
MeV-GeV region. Thus in low-luminosity blazars (mostly BL Lac objects)
X-rays probe the high energy end of the synchrotron component allowing
the investigation of the dynamics of high-energy electrons, while for
powerful blazars (belonging to the FSRQ class) the X-ray emission is
produced via the IC process by relatively low energy electrons.

The study of the X-ray portion of the SEDs of FSRQ is crucial for
different reasons. First of all, the hard X-ray spectrum of FSRQ is
generally attributed to the IC radiation produced by low-energy
electrons scattering ambient photons (External Compton emission). Its
investigation offers the only possibility to probe directly the low
energy portion of the electron distribution which is essential for
estimating the total power of the jet (e.g., Maraschi \& Tavecchio
2003) and is unaccessible at radio frequencies because of
self-absorption. Moreover, the slope of the X-ray continuum, directly
connected to the slope of the underlying population of low-energy
electrons, can provide important constraints on the acceleration
mechanisms acting on the relativistic electrons.  A good description
of the continuum requires broad band observations (0.1-100
keV). Unfortunately, few FSRQs have high statistics above 10 keV, most
of them obtained in the past decade with the PDS instruments onboard
$Beppo$SAX (e.g., Tavecchio et al. 2000, 2002).  The joint fit of the
medium (2-10 keV) and the hard (above 10 keV) X-ray data revealed in
same cases the presence of very hard spectra, with slopes $\alpha
_{\rm hard} \sim 0.3-0.5$ above 2--3 keV (here and in the following
$\alpha $ is defined as $F(\nu)\propto \nu^{-\alpha }$).  These
results have been recently extended using observations with {\it
SWIFT} (Sambruna et al. 2006a, 2007), although in these cases the data
in the hard X-rays represent averages over months.  The presence of
such hard spectra is difficult to reconcile with the standard picture
considering the formation of power law distributions through shock
acceleration or cooling of high energy electrons (e.g., Sikora et
al. 2002).

The medium-soft X-ray spectrum may reveal the very low end of the EC
component and therefore of the relativistic electron spectrum.  Around
these energies a ``break'' is expected in the EC emission which is the
``footprint'' of the thermal photon population. While above the break
the slope of the spectrum is connected to the slope of the electron
distribution, below the break the slope of the spectrum should instead
reveal the spectral distribution of the ambient photons.
Similar ``convex'' spectra have in fact been observed in some
intermediate redshift FSRQs (e.g., Tavecchio et al. 2000, 2002). An
important consequence of the detection of such a break is the
determination of the mimimum Lorentz factor of the emitting electrons,
a key parameter for the determination of the kinetic power carried by
the jet (e.g., Celotti \& Fabian 1993). However, the situation can be
more complex, with the simultaneous presence in this band of different
components, such as the Synchrotron Self-Compton emission and, as
recently pointed out by Celotti, Ghisellini \& Fabian (2007), the spectral
components originating from the ``bulk Comptonization'' of ambient
photons by {\it cold} electrons in the jet.

The interpretation of the shape of the soft X-ray continuum is not
unique. In fact, in several high-redshift blazars ($z>4$; e.g., Yuan
et al. 2006 and references therein) the steepening of the soft X-ray
spectrum is interpreted in terms of absorption by a dense region of
(warm) plasma ($N_H\sim 10^{22}$ cm$^{-2}$) present in the region
surrounding the QSO (expected to be in the form of a wind/outflow,
e.g., Fabian 1999). If confirmed, the presence of this large amount of
absorbing gas in the vicinity of the jet, would have strong
consequences for the structure of the jet and its interaction with
the ISM.


An excellent laboratory to address most of the issues discussed above
is the FSRQs RBS~315 (from the Rosat Bright Survey, Schwope et
al. 2000; also known as GB6~J0225+1846, $z=2.69$), for which a
previous {\it XMM-Newton} observation revealed the presence of an
extremely hard X-ray continuum (photon index $\Gamma\simeq 1.2$, Page
et al. 2005, Piconcelli \& Guainazzi 2005) and a strong deficit of
soft photons. With the aim of obtaining a better description of the
broad band X-ray continuum we asked for a \suzaku observation. In the
same days of the \suzaku observation we obtained a Target of
Opportunity short pointing ($\simeq 4000$ s) with {\it SWIFT}. In the
following we report the data analysis (Sec.2), the results (Sec.3) and
the modelling of the derived SEDs (Sec.4). The discussion is given in
Sec.5, where we explore in particular the possibility that the convex
spectrum reflects the {\it intrinsic} curvature of the IC
component. Conclusions are reported in Sec.6. Throughout this work we
use the following cosmological parameters: $H_{\rm 0}\rm =70\; km\;
s^{-1}\; Mpc^{-1} $, $\Omega _{\Lambda}=0.7$, $\Omega_{\rm M} = 0.3$.

\section{Observations and data reduction}

\subsection{{\it Suzaku}}

The \suzaku payload (Mitsuda et al. 2007) carries four X-ray
telescopes sensitive in the 0.3-12 keV band (XIS, Koyama et al. 2007),
with CCD cameras in the focal plane, together with a non-imaging
instrument (HXD, Takahashi et al. 2007), sensitive in the 10-600 keV
band, composed by a Si-PIN photo-diodes detector (probing the 10-60
keV band) and a GSO scintillator detector (sensitive above 30
keV). Three XIS units (XIS0, 2 and 3) have front-illuminated CCDs,
while XIS1 uses a back-illuminated CCD, more sensitive at low
energies.

\begin{figure}[]
\figurenum{1}
\noindent{\plotone{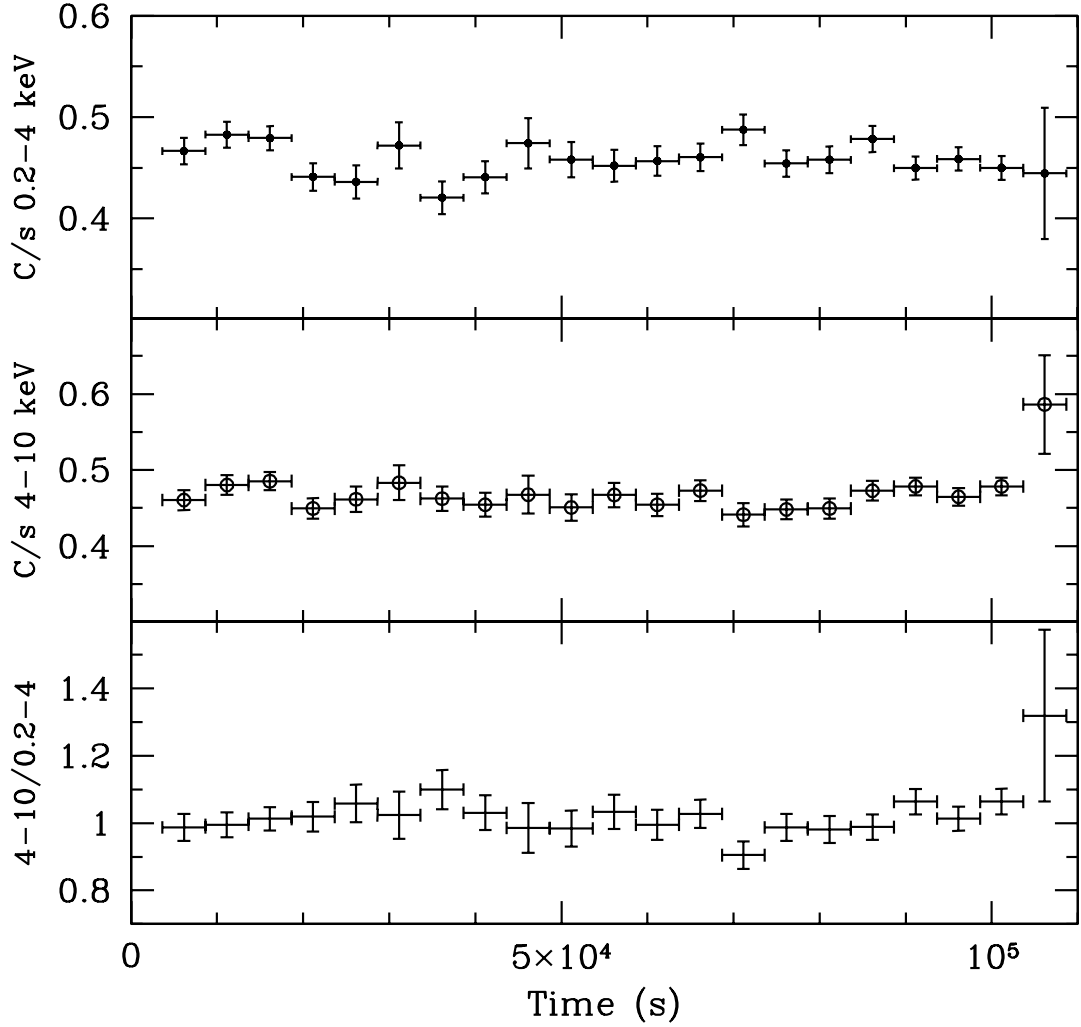}}
\caption{{\it Suzaku}/XIS1 lightcurves of RBS~315 (in bins of 5000 s)
in the soft (0.2-4 keV, upper panel) and hard (4-10 keV, middle panel)
band with the corresponding hardness ratio (bottom panel). No evident
variations are visible within the \suzaku observation.}
\label{lcurve}
\end{figure}

RBS~315 was observed with \suzaku from 2006 July 25 19:36:03 UT to
2006 July 27 00:58:19 UT (sequence number 701077010). The total
on-source time was 49.3 ksec. To increase the sensitivity of the HXD
the observation was performed at the HXD aim-point.
HXD/GSO data are not used in the following analysis, since the
performances and the background of the GSO at the flux level expected
for RBS~315 are still under study.  The analysis have been performed
with the data obtained through the last version of the processing
(v1.2) and the last release of the HEASoft sofware (v6.1.2) and
calibrations. A more extended discussion of the procedure used can be
found in Kataoka et al. (2007). No significant variability is detected
within the observations (see Fig.\ref{lcurve}), therefore spectra were
extracted over the whole exposure.

\subsubsection{XIS data reduction and analysis}

The reduction followed the prescriptions suggested in ``The Suzaku
Data Reduction Guide''\footnote{{\tt
http://suzaku.gsfc.nasa.gov/docs/suzaku/analysis/abc/}; \newline see
also {\tt http://www.astro.isas.ac.jp/suzaku/analysis/}}. Using the
HEASoft tool {\tt xselect} we select good time intervals, excluding
epochs of high background (when the satellite crosses the South
Atlantic Anomaly or the object is too close to the rim of the
Earth). After screening the net exposure time is 35.4 ksec. Events are
then extracted in a circle centered on the source with a radius of
4.6'. Background events are extracted in a similar circle centered in
a region devoid of sources. We checked that the use of
different source and background regions do not significantly affect
the resulting spectra. Response (RMF) and auxiliary (ARF) files are
produced using the tools developed by the \suzaku team ({\tt
xisrmfgen} and {\tt xissimarfgen}) distributed with the last version
of HEASoft. ARFs are already corrected for the degradation of the XIS
response using the tool {\tt xiscontamicalc}.

\subsubsection{HXD/PIN data reduction and analysis}

HXD/PIN data are reduced following the procedure suggested by the
\suzaku team. The HXD/PIN spectrum is extracted after the selection of
good time intervals (analogously to the XIS procedure). To the
extracted spectrum (obtained through {\tt xselect}) we applied the
suggested Dead time correction (of the order of 5\%). The net exposure
time after screening is 40.2 ksec.

\begin{figure}[]
\figurenum{2}
\noindent{\plotone{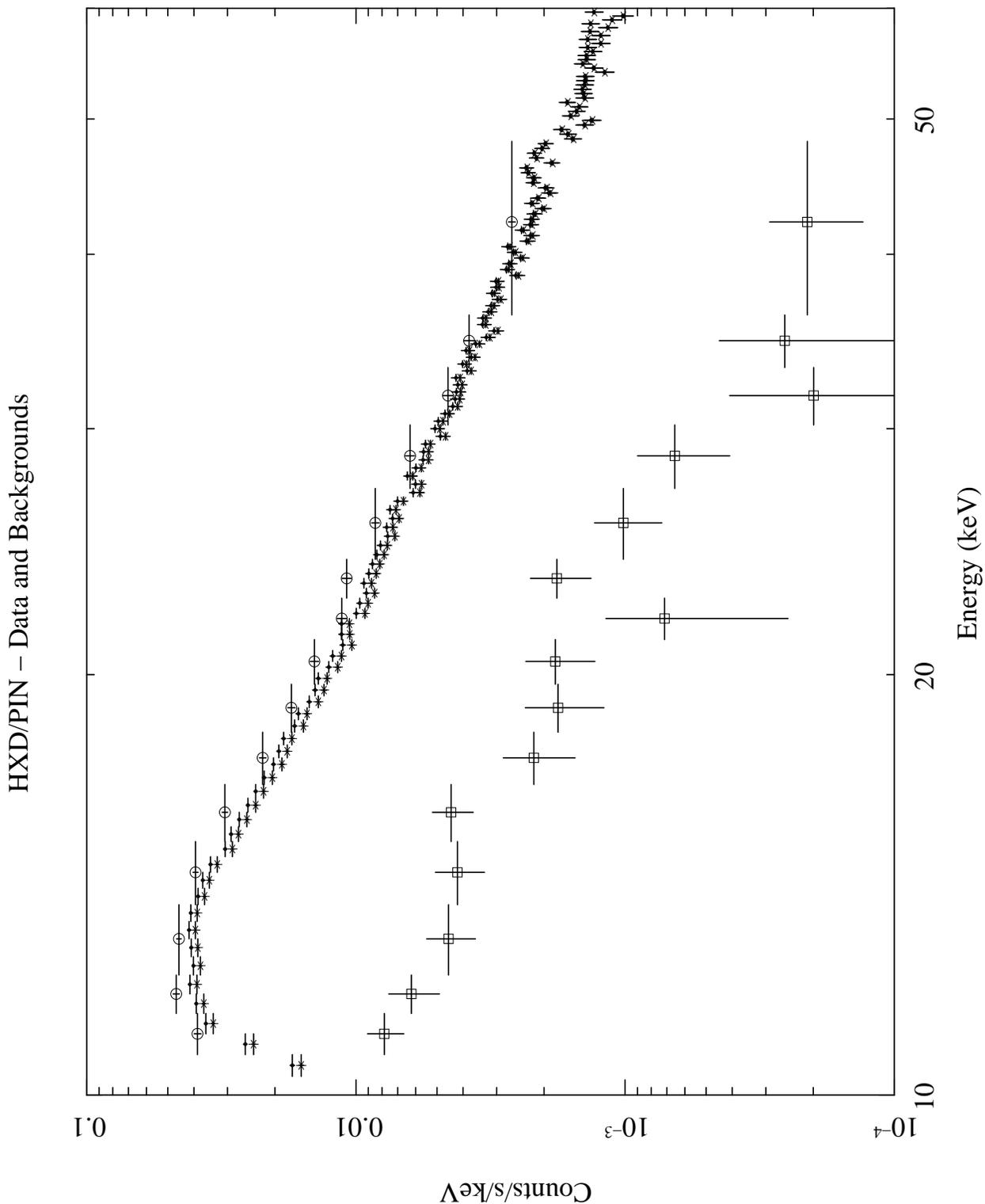}}
\caption{The total HXD/PIN spectrum (open circles) is shown together with the
spectra of the total background (filled circles) and the non-X-ray background
(asterisks). The open squares show the net spectrum of RBS~315.}
\label{bkg}
\end{figure}

Response and non X-ray background (NXB) files are directly provided by the
\suzaku team. For technical reasons, after 2006 May 24 the
response of the PIN has changed, therefore we used response and
background files calculated for this epoch. Note also that, since the
background level of HXD/PIN is extremely low, the background event
files are generated with a ten times scaled level than the actual
background to avoid introducing a large statistical error. The
EXPOSURE keyword in the background file has to be changed before the
analysis. An important issue in the analysis of the HXD/PIN data
concerns the estimate of the Cosmic X-ray Background, whose spectrum
peaks just in this band. We followed the procedure suggested by the
\suzaku team (see also Kataoka et al. 2007), simulating the expected
contribution of the CXB from the entire PIN Field of View ($34'\times
34'$), assuming the {\it HEAO-1} spectrum between 3 and 60 keV (Boldt
1987, Gruber et al. 1999):
\begin{equation}
I_{\rm CXB}(E)=9.0\times 10^{-9} \left(\frac{E}{3\, {\rm keV}}\right)^{-0.29}
\exp \left(- \frac{E}{40\, {\rm keV}}\right) \,\, {\rm erg} \,{\rm cm}^{-2} \,
{\rm s}^{-1} \, {\rm keV}^{-1} \, {\rm sr}^{-1}
\end{equation}
\noindent
It is worth noting that the normalization of the CXB is still under
debate, since the level measured by HEAO-1 in the 2-10 keV band
appears to be below (with a difference of about 20\%) that measured by
more recent missions. However, a recent analysis of the data of the
PDS instrument (in the band 15-50 keV) onboard {\it Beppo}SAX seems to
confirm the HEAO-1 spectrum, but with a slightly larger normalization
(Frontera et al. 2007).  For comparison in Fig.(\ref{bkg}) we report
the total PIN spectrum (open circles), the total background (CXB+NXB,
filled circles), the NXB (asterisks) and the net PIN spectrum
resulting from the background subtraction (open squares). The net
counts represent about 10\% of the total counts. Roughly, the CXB flux
account for 5\% for the HXD/PIN background. We have binned the HXD/PIN
spectra so that each bin contains almost same number of signal
photons\footnote{We rebinned the data with the {\tt grppha} tool using
{\tt group 0 31 2 32 63 4 64 95 8 96 127 32 128 255 64}}. With this
choice each bin has a S/N ratio of about 4.

\subsection{{\it SWIFT}}

{\it SWIFT} (Gehrels et al. 2004) observed RBS~315 two times: on
28/7/2006 from 18:42:40 UT to 20:45:10 UT (for a net exposure of 4152
s) and 31/7/2006 from 15:48:12 UT to 16:21:23 UT (for a net exposure
of 1983). {\it XRT} and {\it UVOT} data have been analized as detailed
in Foschini et al. (2007), using the same version of the software and
calibration. For {\it XRT} we use the data in photon counting mode,
selecting events with grades 0-12.

\subsection{{\it XMM-Newton}}

We reanalized the {\it XMM-Newton} data of the observation of RBS~315
performed on 2003 July 25, previously reported by Page et al. (2005)
and Piconcelli \& Guainazzi (2005). For processing, screening and
analysis of the data we used the same procedure described in Foschini
et al. (2006a), with the latest version (v.7.0.0) of {\tt XMM SAS} and
updated calibration files (20/7/2006). PN, MOS1 and MOS2 instruments
were operated in the full-frame mode. No soft-protons flares occurred
during the exposure: the total exposure time was about 20 ksec on each
detector (18 for PN and 22 for MOS). 

\section{Results}

We fit the data with the {\tt XSPEC} package (v.11.3.2). XIS spectra
were rebinned with a minimum of 50 counts per bin. We checked
that all the XIS units provide consistent results (differences in the
cross-normalizations are of the order of 5\%): therefore we decided to
perform joint fits of all the XIS0,1,2 and 3 spectra. To avoid
problems with the calibration at low energy, we limit the fit to the
range 0.7-10 keV for XIS0,2,3 and to 0.5-10 keV for the XIS1. We also
limit the analysis of the PIN spectrum to the range 12-60 keV (see
Kataoka et al. 2007). All the errors in the parameters are intended at
the 90\% for one parameter of interest ($\Delta \chi^2=2.7$).

\subsection{XIS spectra}

\begin{figure}[]
\figurenum{3}
\noindent{\plotone{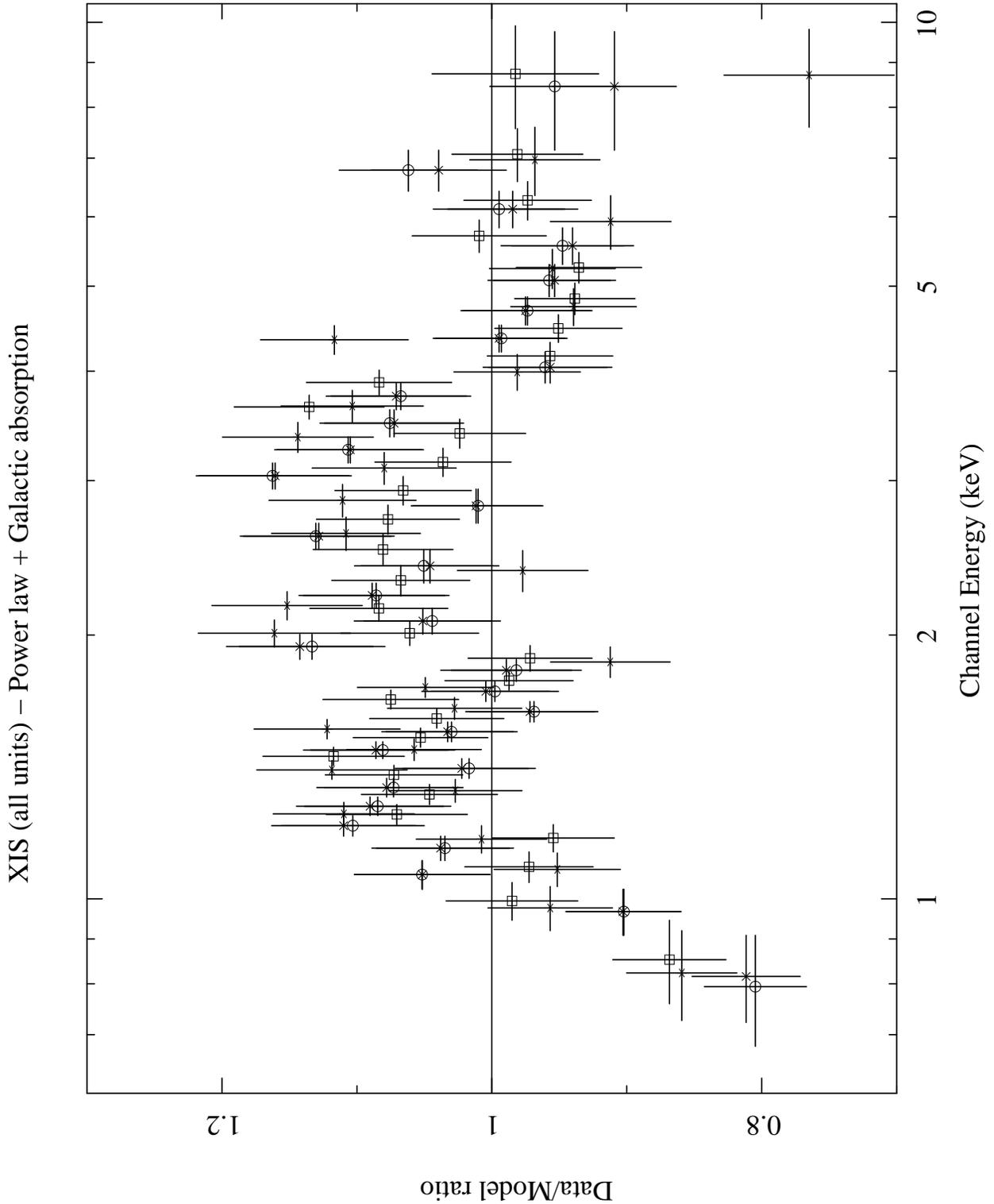}}
\caption{The data/model ratio of the fit of XIS spectra with a
power-law and absorption fixed to the Galactic value. The model is
clearly unadequate to explain the data: in particular there is an
important deficit of soft photons, clearly indicating a more degree of
absorption or an intrinsic hardening of the spectrum below about 1
keV.}
\label{xispowabsgal}
\end{figure}

A power law model with absorption fixed to the Galactic value
($N_H=1.02\times 10^{21}$ cm$^{-2}$, Dickey \& Lockman 1990) clearly
does not provide an adequate fit to the XIS data (Tab.\ref{fitxis},
the data/model ratio is reported in Fig.\ref{xispowabsgal}). If we
allow the absorption to vary we obtain a great improvement of the fit
($\chi^2_r=1.04$), but the value of $N_H$ is clearly in excess to the
Galactic one ($N_H=1.9-2.3\times 10^{21}$ cm$^{-2}$). Then we try to
fit the spectrum assuming Galactic absorption and a broken power law
model. The fit is good ($\chi^2_r=1.043$) and it requires a rather
hard spectrum below 1 keV ($\Gamma=0.68\pm 0.16$). An alternative
possibility is to assume that the intrinsic spectrum is well
reproduced by a power-law with the required extra-absorption provided
by (neutral) absorbing material in the quasar environment. Therefore
we model the data assuming the {\tt zwabs} model in {\tt XSPEC} (plus
the Galactic absorption). Again, the best fit is good, with the value
of the $\chi^2$ ($\chi^2_r=1.047$) almost coincident with that
obtained with the broken power-law model. The value of the required
intrinsic absorption is rather large, $N_{H,z}= 2.6 \times 10^{22}$
cm$^{-2}$ (assuming solar abundance), close to the value found by {\it
XMM-Newton} (Page et al. 2005, Piconcelli \& Guainazzi 2005).  

\subsection{Combined XIS-PIN spectra}

\begin{figure}[]
\figurenum{4}\epsscale{.8}
\plotone{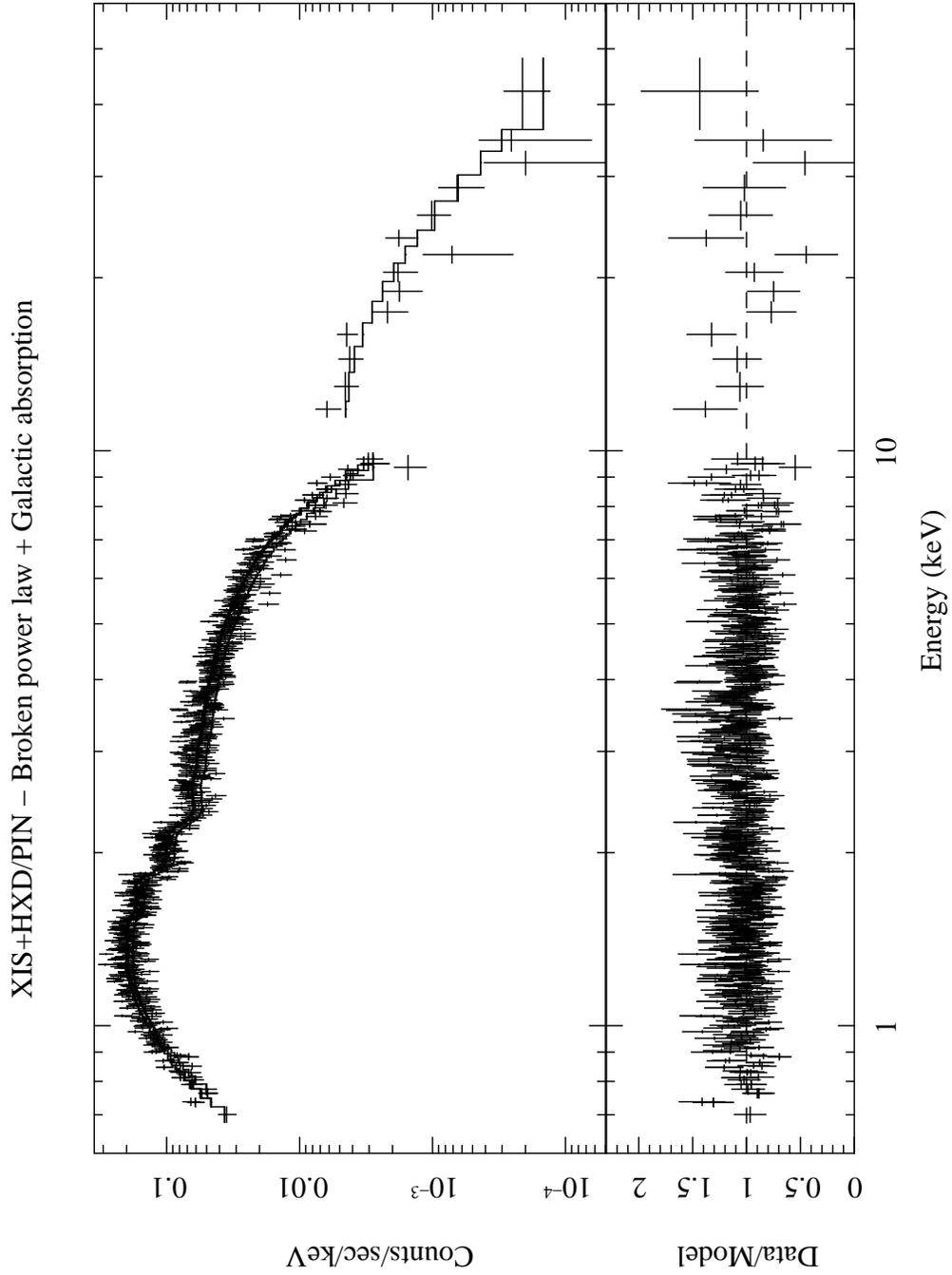}
\caption{Data (upper panel) and data/model ratio (bottom panel) for
the joint fit to the XIS/PIN data with a broken power law and Galactic
absorption.}
\label{xispinbknpowabsgal}
\epsscale{1.0}
\end{figure}

We perform joint XIS and PIN fits (results in Tab.\ref{fitxispin}),
allowing a free normalization factor between the XISs and the PIN to
take into account the uncertainty in the XIS/PIN cross-normalization
(of the order of 15\% at the HXD nominal position\footnote{see {\tt
http://suzaku.gsfc.nasa.gov/docs/suzaku/processing/v1223.html}}). All
the best fit parameters are consistent with those obtained from the
fit of the XIS alone. The required cross normalization factor is
around 20\%, compatible with the instrumental uncertainty. In
Fig.(\ref{xispinbknpowabsgal}) we show the spectrum and the data/model
ratio for the fit with the broken power-law and Galactic absorption.

\subsection{SWIFT/XRT}

Due to the short exposure, the XRT data do not provide strong
constraints on the spectral parameters. Fitting the data obtained
summing the two observations with a power-law absorbed in the
quasar rest frame we derive values consistent with those of \suzaku :
$\Gamma = 1.4 \pm 0.2$, $N_{H,z}=2.4^{+2.1}_{-1.9}\times
10^{22}$ cm$^{-2}$, $F_{2-10}=1.08\times 10^{-11}$ erg cm$^{-2}$
s$^{-1}$ with $\chi^2/d.o.f.=50.9/44$.

\subsection{Comparison with the {\it XMM-Newton} observation}

We fitted the data in the 0.2-10 keV range, after rebinning so that
each energy bin contained a minimum of 50 counts. Results are
presented in Tab.\ref{xmm}.

As already noted by Page et al. (2005) and Piconcelli \& Guainazzi
(2005), a single power-law and Galactic absorption does not provide an
adequate fit to the \xmm data. The fit is satisfactory if we allow
intrinsic absorption or a broken-power law shape for the continuum
(Tab.\ref{xmm}). Consistently with the analysis of Page et al. (2005)
and Piconcelli \& Guainazzi (2005) the broken power law model provides
a worst fit than the absorbed power-law model. In our case, however,
the broken power-law model is still accettable, while Piconcelli \&
Guainazzi (2005) concluded that the broken power law model yelds an
unaccettable fit (with $\chi^2/d.o.f.\simeq 1.2$). The difference with our
analysis is probably due to the refined calibrations used in the more
recent version of the software. Note, moreover, that Piconcelli \&
Guainazzi (2005) used the MOS data in the restricted range 0.8-10 keV,
while the improved calibration allows us to use of the whole range
0.2-10 keV.

\begin{figure}[]
\figurenum{5}\epsscale{.8}
\plotone{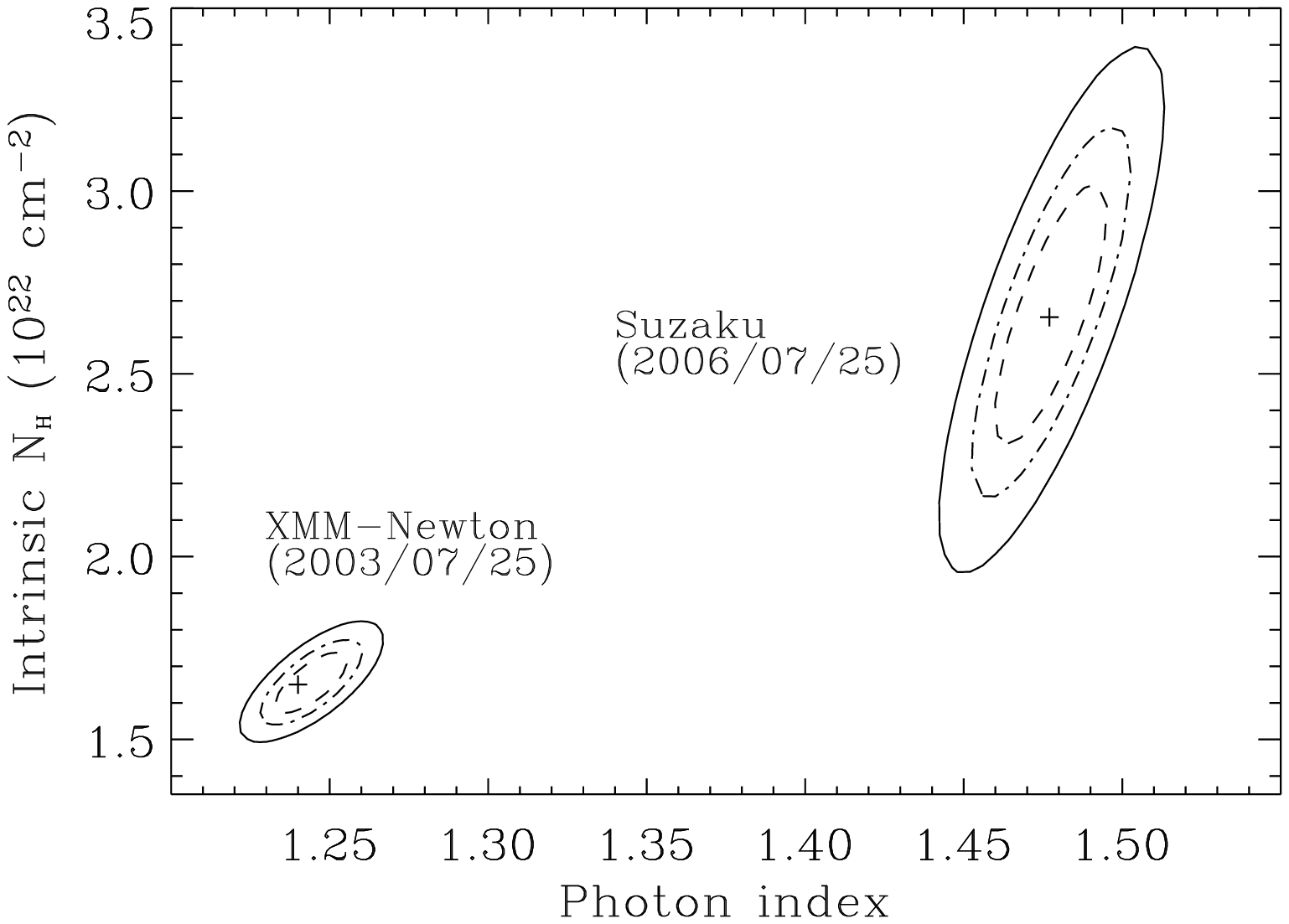}
\plotone{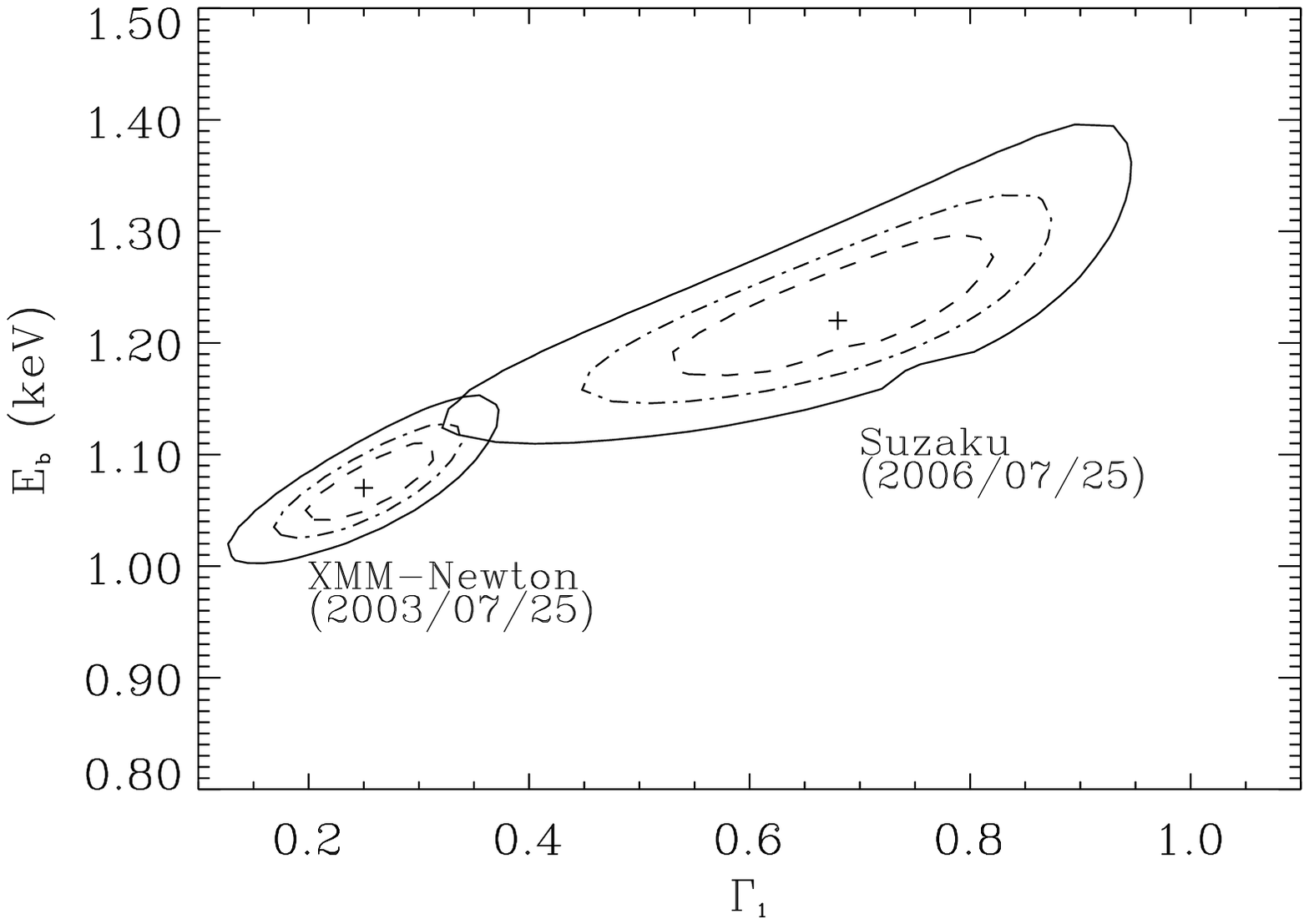}
\caption{{\it Left:} Confidence contours (at 68\%, 90\% and 99\%
levels) for the photon index $\Gamma $ of the power-law and the
intrinsic absorption column density $N_H$, for the fit with the
absorbed power-law model to the {\it XMM-Newton} and {\it Suzaku}
data. The slope of the power-law clearly changed between the two
epochs. {\it Right:} The same for the low-energy photon index and the
break energy for the fit with the broken power-law model.}
\label{contours}
\end{figure}

In Fig.(\ref{contours}) we report the confidence levels (at 68\%, 90\%
and 99\%) for the fit with the power-law absorbed by gas in the quasar
rest frame and for the broken power-law model, calculated for our
\suzaku observation and for the {\it XMM-Newton} data. Although the
required amount of absorbing gas is almost consistent between the two
observations, the slope of the power-law is clearly different in the
two cases. During the \suzaku observation the power law is clearly
softer ($\Gamma \simeq 1.47$) than the previous {\it XMM-Newton}
observation ($\Gamma \simeq 1.26$). On the other hand, the low energy
slope and the break energy of the broken power law are marginally
consistent.

\section{Spectral Energy Distribution and modeling}

\begin{figure}[]
\figurenum{6}
\noindent{\plotone{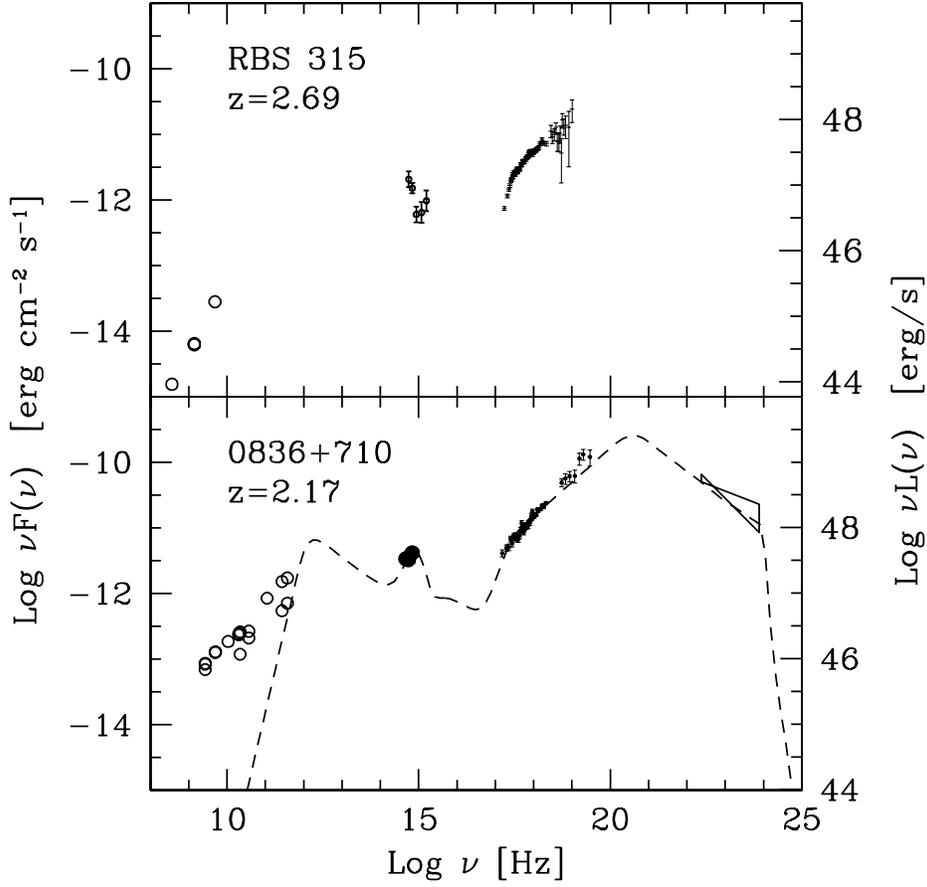}}
\caption{{\it Upper panel:} SED of RBS~315, assembled with archival
radio data (from NED), the optical-UV UVOT data and the {\it Suzaku}
data (for the case of an intrinsically curved continuum, see
text). {\it Lower panel:} for comparison, the SED of 0836+710 (adapted
from Tavecchio et al. 2000), a well studied blazar at comparable
redshift ($z=2.17$). The dashed line is the synchrotron-IC model used
to reproduce the data.}
\label{sedcomp}
\end{figure}

In Fig.(\ref{sedcomp}) (upper panel) we report the SED of RBS~315
constructed using archival radio data from NED, the optical-UV fluxes
obtained with {\it UVOT} and \suzaku data deconvolved for the case of
an intrinsically curved X-ray spectrum. Optical data have been
corrected for Galactic absorption using the maps of Schlegel et
al. 1998 reporting $A_B = 1.193$ mag. If absorption is important in
the quasar rest frame, the points should be considered as lower limits
to the real flux.

For comparison, in the lower panel we report the SED of 0836+710 (from
Tavecchio et al. 2000), a well known $\gamma$-ray loud blazar (also
characterized by an extremely flat X-ray continuum and showing
indications of absorption/hardening at low energy), located at
comparable distance ($z=2.17$). The similarity between the two SEDs is
remarkable, strongly supporting the classification of RBS~315 as a
blazar. Both sources lies at the top of the so-called blazar sequence
(Fossati et al. 1998) with apparent luminosities in excess of
$10^{48}$ erg/s. The comparison also suggests that RBS~315 should be
quite bright in the $\gamma $-ray band, although it was not not
detected by {\it EGRET/CGRO}.
We expect that {\it GLAST}, with its wide field of view,
allowing the complete monitoring of the sky in few hours and higher
sensitivity, should easily detect RBS~315.
 
\begin{figure}[]
\figurenum{7}
\noindent{\plotone{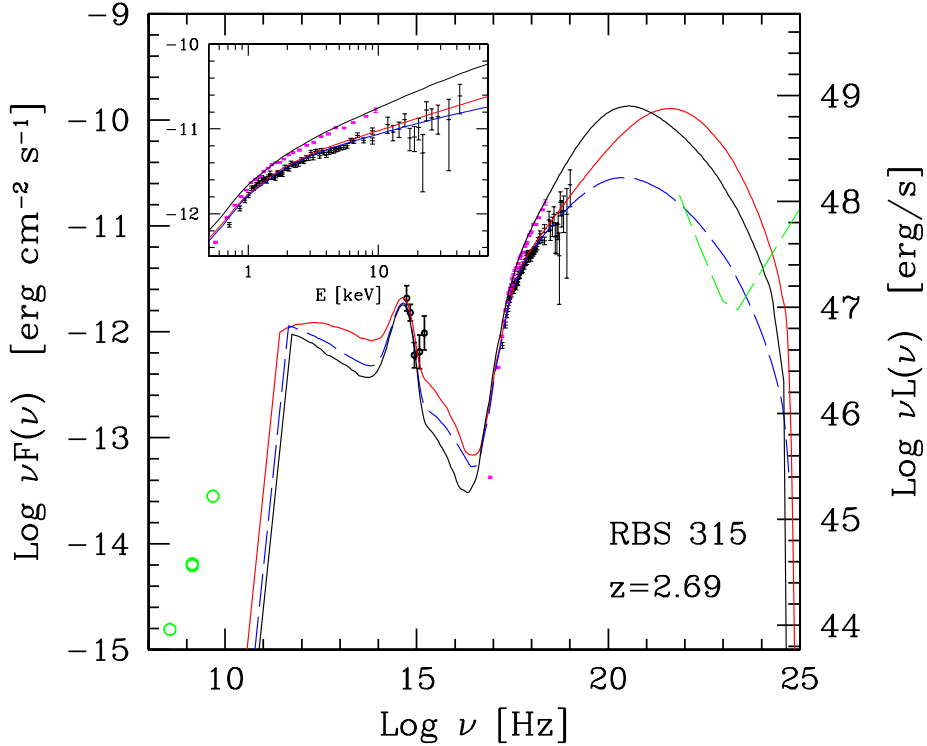}}
\caption{Spectral Energy Distribution of RBS~315 constructed with
historical radio data (from NED), optical-UV data obtained with
UVOT/SWIFT and X-ray data from {\it XMM-Newton} and \suzaku . We
report X-ray data corresponding to the case of an intrinsically curved
continuum. The insert shows a zoom on the X-ray band. The three curves
correspond to the models used to reproduce the data. See text for more
details. The bump in the optical-UV region represents the approximated
disk emission. The long-dashed line reports the {\it GLAST} 5$\sigma$
sensitivity for 1 year.}
\label{sedall}
\end{figure}

\begin{figure}[]
\figurenum{8}
\noindent{\plotone{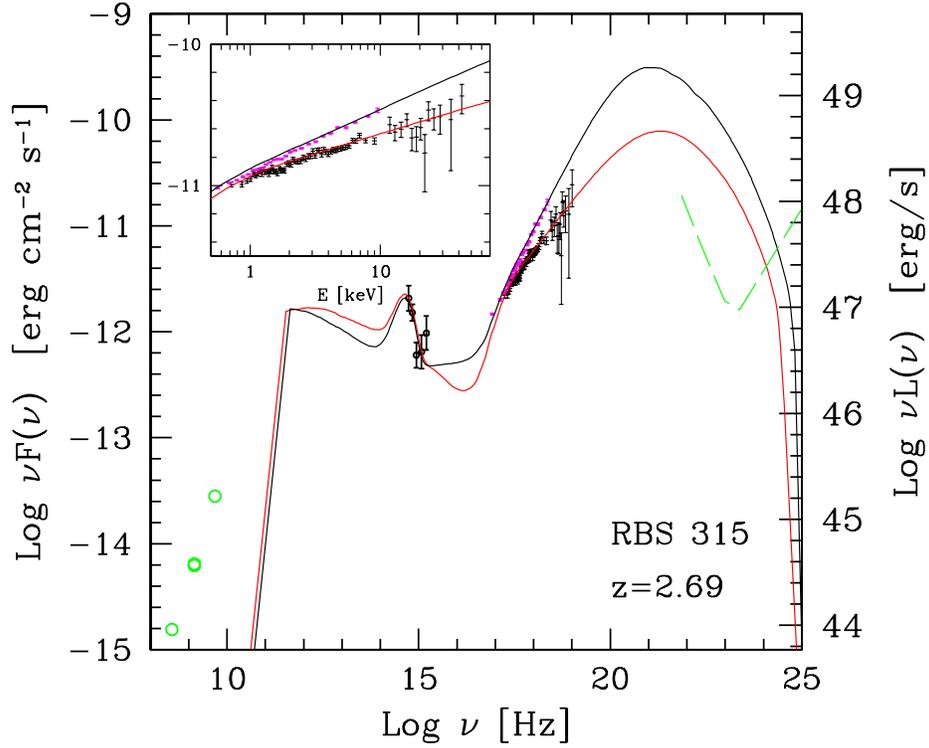}}
\caption{Spectral Energy Distribution of RBS~315, as in
Fig.(\ref{sedall}), but with X-ray data corresponding to the case of
absorption in the quasar rest frame. To reproduce the X-ray continuum
it is necessary to decrease the frequency of the intrinsic break of
the IC continuum, by decreasing the bulk Lorentz factor of the flow.}
\label{sedallabs}
\end{figure}

Given the ambiguity on the shape of the intrinsic X-ray continuum, in
the following we will discuss the SEDs for RBS~315 separately for the
case of intrinsic curvature ({\it Curved, C}), and for the absorbed
power law ({\it Absorbed, A}). The X-ray data, unfolded with the two
models, are shown in Figs.(\ref{sedall})-(\ref{sedallabs}). In the
second case the intrinsic X-ray spectrum emitted by the jet is a
straight power-law
We consider both the X-ray state corresponding to the {\it XMM-Newton}
observation in 2003 and that observed by \suzaku in 2006.

The optical-UV points probably trace the so-called ``blue bump'',
i.e. the direct emission from the accretion disk, whose luminosity
is then rather large, $L_D=4\times 10^{47}$ erg/s.

We reproduced the SED with the standard model for blazars, assuming
synchrotron and IC (both SSC and EC) emission from relativistic
electrons in the jet. Specifically (see Maraschi \& Tavecchio 2003 for
a full description) we assume a spherical region with radius $R$,
moving with bulk Lorentz factor $\Gamma _b $ at an angle of view
$\theta \simeq 1/\Gamma _b$ with respect to the line of sight (under
these conditions the relativistic Doppler factor is $\delta =\Gamma
_b$). The region is filled by a tangled magnetic field $B$ and by
relativistic electrons, assumed to follow a smoothed broken power-law
energy distribution between the Lorentz factors $\gamma _{\rm min}$
and $\gamma _{\rm max}$, with slopes $n_1$ and $n_2$ before and after
the break at $\gamma _{\rm p}$. This {\it purely phenomenological}
form has been assumed to reproduce the observed shape of the blazar
SEDs, without any specific assumption on the acceleration/cooling
mechanism acting on the particles. With this choice we are allowed to
assume extreme low-energy slopes ($n_1<2$) such as those required for
RBS~315 in 2003, which cannot be obtained under standard
conditions. If the radiative cooling timescale is shorter than the
light crossing time of the source we expect $n_1\geq 2$. We remind
that the model cannot reproduce the emission in the radio band since
the source is opaque in this band due to synchrotron
self-absorption. The radio emission is generally assumed to originate
from larger regions of the jet, downstream to the blazar region.

The IC process involves both the synchrotron photons (Synchrotron
self-Compton, Maraschi et al. 1992) and ambient photons (from the disk
[Dermer \& Schlickeiser 1993] and/or reprocessed by the Broad Line
Region [Sikora et al. 1994]) entering the jet (External Compton). In
sources like RBS~315 the EC emission dominate the high-energy
component, while the SSC component could contribute in the soft X-ray
band.

In the case in which the ambient photons are dominated by the emission
from the BLR clouds it is customary to model the external photon
spectrum with a black body shape peaking around $\nu _{BB}\sim
10^{15}$ Hz (e.g., Ghisellini et al. 1998). Although rather crude, the
IC spectra derived from this approximation reproduces quite well more
refined spectra calculated taking into account the exact shape of the
BLR spectrum (Tavecchio et al., in prep.). To calculate the EC spectra
we have also to assume the energy density of the BLR radiation,
$U_{BLR}$, given by: $U_{BLR}\simeq L_{BLR}/4\pi c
R^2_{BLR}$. Unfortunately we could not found useful information on
$L_{BLR}$, the luminosity of the emission lines of this
source. Therefore we indirectly fix $L_{BLR}$ assuming that optical-UV
emission directly measure the disk luminosity, $L_D$, and that a
fraction $\tau = L_{BLR}/L_D=0.1$ of the disk emission is reprocessed
by the BLR (this is consistent with the results obtained for the few
sources in which a measure of the disk and line luminosity is
simultaneously available, e.g., Sambruna et al. 2006b). The
luminosity-radius relations for the BLR (Kaspi et al. 2005, Bentz et
al. 2006) extrapolated at these large luminosities would imply quite
large BLRs, with radii $R_{BLR}> 10^{19}$ cm. These large radii would
imply small $U_{BLR}$. However, the validity of these relations in
these extreme regimes is not tested. Moreover, besides the BLR
radiation, we may have the radiation coming directly from the
accretion disk and/or isotropized by intercloud scattering
material. Therefore we assume $R_{BLR}=10^{18}$ cm, which implies a
relatively large radiation energy density.

The models are reported in Figs.(\ref{sedall})-(\ref{sedallabs}) and
the model parameters are reported in Tab.(\ref{parameters}), together
with the derived jet radiative luminosity and jet power (see the
Discussion). We report the models calculated both in the case of the
X-ray spectrum measured by {\it XMM-Newton} in 2003 and that derived
by \suzaku in 2006.

In Fig. (\ref{sedall}) we report the models for the {\it C} case (the
insert shows a zoom on the X-ray band). The high-energy emission, from
X-rays to $\gamma $-rays, is dominated by the EC component. This can
be approximately described by three power laws: 1) an extremely hard
power law below a ``break frequency'' $\nu _{br,obs}\simeq \nu _{BB}
\Gamma _b ^2 \gamma _{\rm min }^2/(1+z)\sim 10^{17}$ Hz (where $\nu
_{BB}\simeq 10^{15}$ Hz, $\Gamma _b =20$, $\gamma _{\rm min}=1$,
values typically derived in these sources, e.g., Tavecchio et
al. 2000); 2) a power law with spectral index $\alpha _1=(n_1-1)/2$
between $\nu _{br,obs}$ and $\nu _{p,obs}\simeq \nu _{BB} \Gamma _b ^2
\gamma _{\rm p}^2/(1+z)$ (marking the position of the EC peak); 3) a
steeper power law with slope $\alpha _2=(n_2-1)/2$ up to the frequency
$\nu _{max,obs}\simeq \nu _{BB} \Gamma _b ^2 \gamma _{\rm
max}^2/(1+z)$. We remark that the steepening of the spectrum below
$\nu _{br,obs}$ is {\it required} by the standard emission models for
this kind of sources, although the precise shape and position of the
break depends on the jet parameters and on the assumed underlying
ambient photon population (Tavecchio et al., in prep). In particular,
the slope below $\nu _{br,obs}$ reflects the shape of the underlying
soft photons. Using the black-body approximation the resulting
spectrum is $F(\nu)\propto \nu^2$. The slope of the power law above
$\nu _{br,obs}$, $\alpha _1$, is fixed by the measured slope of the
power-law derived from the fitting of the X-ray data.  Note that the
presence of a strong depression between the UV-optical and the
soft-X-ray band requires a minor contribution from the SSC component,
which peaks just in this region.

Without any information on the $\gamma$-ray emission, the values
of $\gamma _{\rm p}$ and $\alpha _2$ are basically unconstrained.
For illustration, we report two realizations of the model reproducing
the \suzaku X-ray spectrum with two different values of $\gamma _{\rm
p}$, the Lorentz factor of the electrons emitting at the peak ($\gamma
_{\rm p}=100$, solid, and $\gamma _{\rm p}=20$, long dashed). Clearly,
the two cases mainly differ for the expected power output in the
$\gamma $-ray band, the first being above and the second below the
{\it EGRET} limit of sensitivity (around $10^{-11}$ erg cm$^{-2}$
s$^{-1}$, e.g., Thompson et al. 1993). Only simultaneous $\gamma $-ray
observations, providing a constraints on the position of the EC peak,
can allow us to fix the value of $\gamma _{\rm p}$. For comparison we
report the predicted {\it GLAST} 5$\sigma$ sensitivity (long-dashed
curve\footnote{\tt
http://www-glast.slac.stanford.edu/software/IS/glast\_lat\_performance.htm})
calculated for 1 year all-sky survey.

The models for the case of intrinsic absorption are reported in
Fig.(\ref{sedallabs}). In this case the X-ray spectrum of the source
is a power-law extending down to lower energy. To correctly reproduce
this spectrum we decrease the bulk Lorentz factor with respect to the
previous case (to $\Gamma _b =13$), so that the hard tail falls
outside the region probed by the data. We compensated the lower degree
of beaming with a larger electron density, with the effect to increase
the relative importance of the SSC emission, which fills the region
between the optical-UV and the soft X-ray band.

In both cases the \suzaku X-ray spectrum can be well fitted by
assuming a standard slope $n_1=2$ for the low energy branch of the
electron energy distribution. On the other hand, the extremely hard
X-ray continuum measured by {\it XMM-Newton} requires a rather flat
distribution ($n_1=1.5$). Apart for the change of $n_1$, the
transition between the state of 2003 to that of 2006 can be reproduced
with small changes in the other parameters, in particular a slightly
larger value of the magnetic field.

\section{Discussion}

\subsection{Absorption or intrinsic curvature?}

Both the \xmm and the \suzaku data of RBS~315 show a clear hardening
of the X-ray spectrum below 1 keV, equally well fitted by a power-law
model with absorption in excess to the Galactic value or by a
broken-power law continuum with Galactic absorption.  This feature,
observed in several high-$z$ quasars, is commonly ascribed to the
presence of absorbing material in the quasar environment (e.g., Elvis
et al. 1994, Cappi et al. 1997, Reeves et al. 1997, Fiore et al. 1998,
Fabian et al. 2001a,b, Bassett et al. 2004, Sambruna et al. 2007).  A
debated issue is the possible existence of a trend for the absorption
to increase with redshift (Elvis et al. 1994, Cappi et al. 1997, Fiore
et al. 1998, Page et al. 2005, Yuan et al. 2006). This would support
scenarios (e.g. Silk \& Rees 1998, Fabian et al. 1999) proposing that
in the earliest phases of their evolution QSOs are substantially
obscured by gas, subsequently expelled from the host galaxy by
powerful winds.

The observational evidence, however, is still far from being
completely clear. For instance, it seems that absorption is rather
more common in radio-loud than in radio-quiet quasars (e.g. Page et
al. 2005, Grupe et al. 2006), suggesting that absorption is linked to
the presence of a relativistic jet. However, as pointed out by Page
et al. (2005), if the $N_H-z$ correlation is real, the fact that
radio-loud sources appear to be more absorbed could simply be due to a
selection effect, since radio-loud quasars are brighter in X-rays and
can be observed at larger distances. We note that a substantial
number of the high $z$ radio loud quasars are in fact blazars,
i.e. observed at small angle to the jet axis as is the case for
RBS~315 and the very hard X-ray spectra indicate that their emission
is dominated by the jet.

A potentially severe problem of the absorption interpretation is the
apparent absence of strong extinction in the optical-UV band. Indeed,
with the column densities inferred from the X-ray spectral fits, in
some cases reaching values of $N_H=10^{23}$ cm$^{-2}$ (Fabian et
al. 2001a), the corresponding obscuration would be huge ($A_V\sim
100$), in contrast with the optical observations requiring rather
small reddening.  A solution to this discrepancy is to invoke an
extreme gas to dust ratio in these sources, possibly due to the
high-ionization state of the gas (as the warm absorber found in
Seyfert galaxies). However, the evidence of the high-ionization state
from the fits of the X-ray spectra is not conclusive, even with good
quality data (e.g. Worsley et al. 2004a,b). In fact, due to the
redshift, the most important features that can be used for the
diagnostics move below 0.1 keV and cannot be observed.

\begin{figure}[]
\figurenum{9}
\noindent{\plotone{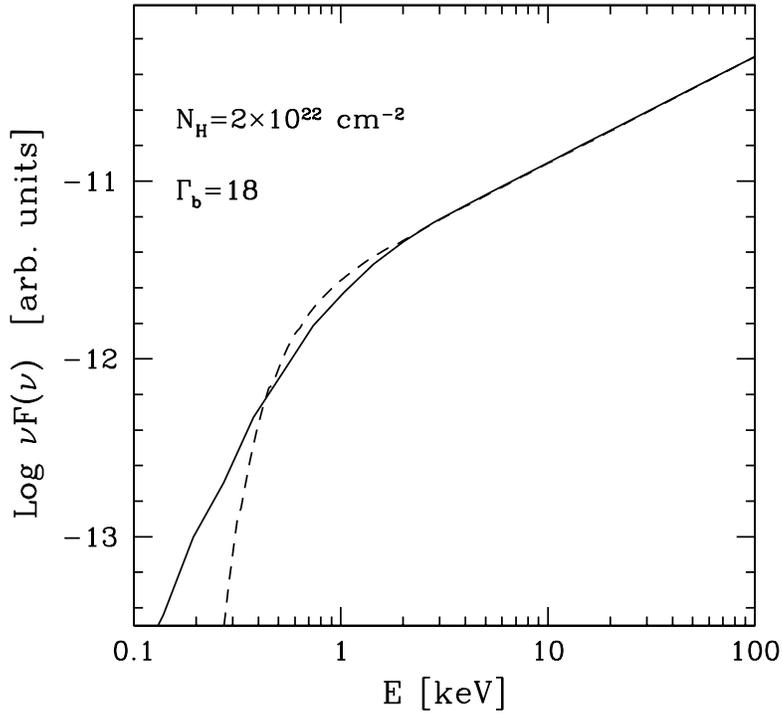}}
\caption{Comparison between the spectrum resulting from a power-law
(with index $\alpha _X=0.4$) absorbed by material with column density
$N_H=2\times 10^{22}$ cm$^{-2}$ in the quasar rest frame (dashed line)
and the EC spectrum (solid line). We assume $\gamma _{\rm min}=1$,
while the bulk Lorentz factor $\Gamma _b=18$ is fixed to closely
reproduced the absorbed power-law spectrum. The source is assumed
at $z=3$ and the spectra have been arbitrarily normalized.}
\label{compa}
\end{figure}

The discussion above is based on the assumption that the hardening at
low energies is due to (intrinsic) absorption in both radio-quiet and
radio-loud sources. However in radio-loud sources, and in particular
in blazars, the observed spectral shape could be {\it naturally}
accounted for by the intrinsic curvature of the IC emission from the
jet, as shown here for the case of RBS~315 (see also Fabian et
al. 2001a; Sambruna et al. 2006a, 2007). Indeed, for parameters
typically derived in reproducing the SED of powerful blazars, the
model predicts a smooth hardening of the spectrum below a few keV (in
the rest frame). To better illustrate this point, we show in
Fig.(\ref{compa}) the comparison between the EC spectrum (solid line)
and an absorbed power law spectrum (dashed) for an intrinsic column
density $N_H=2\times 10^{22}$ cm$^{-2}$ (calculated using the {\tt
zwabs} model of {\tt XSPEC}), for a source located at $z=3$ (energies
are in the observer frame). As already discussed, the EC spectrum
present a break at a frequency approximately given by $\nu
_{br,obs}\simeq \nu _{BB} \Gamma _b ^2 \gamma _{\rm min }^2/(1+z)$,
where $\nu _{BB}$ is the peak frequency of the external radiation
field. To mimic an effective column density one has therefore to
adjust the parameters to obtain the right value for the break
frequency. For instance, the curve reported in Fig.(\ref{compa}) have
been calculated assuming $\nu _{BB}=10^{15}$ Hz and fixing the
low-energy end of the electron distribution to $\gamma _{\rm
min}=1$. Therefore, $\nu _{br,obs}$ is fixed only by the value of the
bulk Lorentz factor.  As can be seen, the shapes of the two kinds of
spectra are substantially similar down to energies of 0.3-0.4 keV (but
this energy is lower for sources at higher $z$). Below that energy,
the EC curve stays systematically above the absorbed power-law, since
it asymptotically reaches a spectrum $\propto \nu^2$ (reproducing the
slope of the underlying population of soft photons, assumed to follow
a black-body shape), while the absorption imprints an exponential
cut-off. However, with data covering a limited energy range the two
models can give equally acceptable fits.


The hardening of the EC spectrum at low energies could be masked by
the possible presence of the softer SSC component, which typically
peaks between the UV and the soft X-ray band. However, from the
observational point of view (e.g. Fossati et al. 1998) the relative
importance of the SSC component seems to decrease with the power of
the source (maybe due to an increasing importance of the accretion
disk and then of the external radiation field: e.g., Celotti et
al. 2007), leaving an almost ``naked'' EC component in the most
powerful sources. In this case, the apparent effect of the absorption
should be more important for the most powerful sources, accounting for
the apparent trend of the $N_H$ with $z$ (since for high redshift we
select the most powerful sources).

In conclusion, in radio-loud, blazar-like, quasars, the intrinsic
curvature of the EC component is a viable alternative to explain the
cut-off in the soft X-ray band.
Clearly, more observational effort should be devoted to study this
effect. The most direct observation able to discriminate between the
two possible explanations would be the direct detection of features
(i.e. absorption edges) directly imprinted on the X-ray continuum by
the absorbing material. Spectra with sufficient signal can possibly be
obtained with long exposures with the present instrumentation.
In any case, this can be one of the goals of the next generation of
X-ray instruments. Moreover, it is mandatory to investigate the
detailed shape of soft the X-ray continuum expected by considering
realistic spectra for the soft ambient photons scattered in the EC
component (Tavecchio et al., in prep).

\subsection{Flat X-ray continuum and variability}

Besides the problems related to the cut-off at low energies, the quite
hard X-ray spectrum of RBS~315 above 1 keV (similar to that of other
powerful blazars) poses a great challenge to the scenarios currently
accepted for particle acceleration (Ghisellini 1996, Sikora et
al. 2002). Indeed, to correctly reproduce the X-ray continuum we have
to assume $n_1<2$ (especially for the \xmm observations). However, a
slope $n=2$ for the electron energy distribution (implying a spectral
index $\alpha _X=0.5$) is predicted in scenarios considering either
shock accelerated electrons or a population of cooled electrons.
Particularly difficult to be reconciled with the standard models is
the exceptionally hard slope measured by \xmm in 2003. Note, moreover,
that such hard spectra are not uncommon. For instance, in the sample
of 16 radio-loud quasars at $z>2$ considered by Page et al. (2005), 4 have
hard spectra with $\Gamma <1.4$.

This evidence could suggest either that shocks can produce
relativistic electrons with a distribution much harder than $n=2$ or
that another mechanism energizes the electrons, at least those with
Lorentz factors below $\gamma \sim 10$ (which typically emits through
EC at energies below 100 keV). This possibility has been discussed by
Sikora et al. (2002) which proposed a two-step acceleration
process. In this scenario, electrons are initially accelerated from
thermal energies to relativistic ones by a stochastic process
(involving, for instance plasma instabilities/turbulence or following
magnetic reconnection events), thought to provide electron
distributions harder than $n=2$. Only those particles with energies
above a threshold can effectively be accelerated by the Fermi process
at the shock front at energies at which they can emit $\gamma $-rays.
However these speculations need to be assessed by improved theoretical
analysis and calculations.

Another interesting point concerns the variability. Although extremely
variable in the $\gamma $-rays, FSRQ show little variability in the
X-ray band on small timescales (days). On longer timescales
($>$months) variability is observed, but the variations are usually
small, both in flux and slope (see, for instance, the comparison
reported in Sambruna et al. 2007). Only in few cases extreme
variability has been observed both on long (e.g., Pian et al. 2006)
and short (e.g., Foschini et al. 2006b) timescales. RBS~315 changes
the slope of the X-ray spectrum from $\Gamma =1.3$ to 1.5 between the
two observations, separated by 3 years. An interesting point to note
is that the spectrum seems to change ``pivoting'' around 1 keV (see
Fig.\ref{sedall},\ref{sedallabs}). Sources displaying a behaviour
similar to RBS~315 are RX J1028.6-0844 ($z=4.276$), for which two \xmm
observations separated by one year show a change of the photon index
from $\Gamma \simeq 1.3$ to 1.5 (Yuan et al. 2005), and 0836+710, for
which the hard X-ray spectrum (20-200 keV) softens from $\Gamma\simeq
1.4$ (as measured by {\it BeppoSAX}) to $\Gamma \simeq 1.8$ ({\it
BAT/SWIFT}) (Sambruna et al. 2007).

\subsection{Jet power and jet/disk connection}

In Tab.{\ref{parameters}} we also report the radiative luminosity,
$L_{\rm rad}$, and the power carried by the jet, $P_j$, calculated
assuming the presence of 1 proton for each relativistic electron. This
choice is dictated by the condition that the jet carries enough power
to support the large radiative output of these sources (e.g. Maraschi
\& Tavecchio 2003). The derived values are around $P_j\sim 10^{48}$
erg/s, quite large, but consistent with the values inferred for
sources of similar radiative luminosity (e.g. Tavecchio et
al. 2000). The ratio $L_{\rm rad}/P_j$, indicating the radiative
efficiency of the jet, spans the range 10-50, again, typical for this
kind of sources (e.g. Sambruna et al. 2006b).

In the hypothesis that the disk substantially contributes to the
optical-UV luminosity, one can derive the disk luminosity,
$L_D=4\times 10^{47}$ erg/s. As already discussed for other powerful
blazars, in the case of the standard value of $\sim 10\%$ for the
accretion efficiency, this would imply that the jet carries away a
power comparable to that carried inward by the accretion flow
(Rawlings \& Saunders 1991, Celotti et al. 1997, Maraschi \& Tavecchio
2003). Finally we note that even if the black hole is accreting at the
Eddington limit, the large luminosity requires an extreme value for
its mass, $M_{BH}=3\times 10^9$ $M_{\odot}$.

\section{Conclusions}

\noindent
$\bullet $ The analysis of the \suzaku spectrum of RBS~315 confirms
the finding of Page et al. (2005) and Piconcelli \& Guainazzi (2005)
made using \xmm data on the presence of a cut-off in the soft-X-ray
spectrum. This can be fitted equally well either using a power-law with
absorption in excess of the Galactic value or a broken power-law with
Galactic absorption.

\noindent
$\bullet $ The very hard spectrum measured by \xmm (photon index
$\Gamma \sim 1.2$) poses a difficult challenge to the models usually
considered for the acceleration of relativistic electrons. During the
\suzaku observation, however, the slope was steeper, $\Gamma =1.5$,
closer to more ``standard'' values.

\noindent
$\bullet $ The SED, constructed with historical radio data, optical-UV
data obtained with {\it UVOT} and the X-ray data, is consistent with
those usually derived for other intermediate/high-$z$ powerful
blazars, usually shining in the $\gamma $-ray band. It is thus likely
that RBS~315 will be detected by the upcoming {\it GLAST} satellite.

\noindent
$\bullet $ Through the modeling of the SED with the standard
synchrotron+IC emission model for blazars we derive parameters typical
for powerful blazars. However, the lack of the $\gamma $-ray detection
prevent a firm constraint of all the parameters.

\noindent
$\bullet $ We showed that the cut-off in the soft X-ray band, commonly
ascribed to absorption by material in the quasar environment, can be
also naturally reproduced by the hard tail of the IC
component. Further studies are required to distinguish between the two
alternatives.

\acknowledgments We thank Giancarlo Ghirlanda for useful discussions
and the anonymous referee for constructive comments. We are grateful
to Niel Gehrels and the {\it Swift } team for the ToO observation of
RBS~315. We acknowledge financial support from ASI. This research has
made use of the NASA/IPAC Extragalactic Database (NED) which is
operated by the Jet Propulsion Laboratory, California Institute of
Technology, under contract with the National Aeronautics and Space
Administration.


 
 
 
 
\newpage 


\begin{table}[h]
\begin{center}
\begin{tabular}{lccccc}
\multicolumn{6}{c}{XIS}\\ \hline 
Model & $\Gamma/\Gamma _1$ & $N_{H,z}/\Gamma_2$ & $E_b$ & $\chi ^2/d.o.f.$ & $F_{2-10}$ \\ 
(1)&(2)&(3)&(4)&(5)&(6)\\\hline 
&&&&&\\
pl+GA& $1.37 \pm 0.01$& --& -- & 1252.9/1061& 8.5$\pm 0.1$\\ 
pl+GA+zA& $1.48\pm 0.02$ & 2.65$\pm 0.4$& -- & 1110.4/1060& 8.4$\pm 0.1$\\ 
bpl+GA& 0.68$\pm 0.16$& 1.42$^{+0.02}_{-0.01}$& 1.23$\pm 0.08$& 1105.2/1059& 8.4$\pm 0.1$\\ 
&&&&&\\ \hline
\end{tabular}
\end{center}
\caption{\scriptsize Best fit parameters for the XIS data. Description
of columns: (1): Model used to fit the data (pl=power law; bpl=broken
power law; GA=Galactic absorption, {\tt wabs}; zA: neutral absorption
in the quasar rest frame, {\tt zwabs}). (2) Photon index for the pl
model or low-energy photon index for the bpl model. (3) Value of the
intrinsic $N_H$ (in units of $10^{22}$ cm$^{-2}$) for the zA model or
high-energy photon index for the bpl model. (4) Break energy (keV) for
the bpl model. (6) Flux in the 2-10 keV band, in units of $10^{-12}$
erg cm$^{-2}$ s$^{-1}$.}
\label{fitxis}
\end{table}

\begin{table}[h]
\begin{center}
\begin{tabular}{lccccccc}
\multicolumn{8}{c}{XIS+HXD/PIN}\\\hline
Model & $\Gamma/\Gamma _1$ & $N_{H,z}/\Gamma_2$ & $E_b$ & $N_{\rm PIN/XIS}$&$\chi ^2/d.o.f.$ &$F_{2-10}$& $F_{10-50}$\\ 
(1)&(2)&(3)&(4)&(5)&(6)&(7)&(8)\\\hline 
&&&&&&&\\
pl+GA+zA& $1.48\pm 0.02$  &  2.7$\pm 0.4$ &--  & $1.28\pm 0.17$  & 1123.0/1073   &  8.3$\pm 0.1$  & 25$\pm2$\\ 
bpl+GA &0.68$\pm 0.15$ & 1.42$^{+0.02}_{-0.01}$ &1.22$^{+0.08}_{-0.06}$& $1.19\pm 0.15$& 1118.1/1072  & 8.4$\pm 0.1$ & 25$\pm2$\\ 
&&&&&&&\\\hline
\end{tabular}
\end{center}
\caption{\scriptsize Best fit parameters for the XIS+HXD/PIN
data. Description of columns: (1): Model used to fit the data
(pl=power law; bpl=broken power law; GA=Galactic absorption, {\tt
wabs}; zA: neutral absorption in the quasar rest frame, {\tt zwabs}). (2)
Photon index for the pl model or low-energy photon index for the bpl
model. (3) Value of the intrinsic $N_H$ (in units of $10^{22}$
cm$^{-2}$) for the zA model or high-energy photon index for the bpl
model. (4) Break energy (keV) for the bpl model. (5) Value of the
PIN/XIS cross normalization factor. (7) Flux in the 2-10 keV band, in
units of $10^{-12}$ erg cm$^{-2}$ s$^{-1}$. (8) Flux in the 10-50 keV
band, in units of $10^{-12}$ erg cm$^{-2}$ s$^{-1}$.}
\label{fitxispin}
\end{table}

\newpage

\begin{table}[h]
\begin{center}
\begin{tabular}{lccccc}
\multicolumn{6}{c}{{\it XMM-Newton}}\\ \hline 
Model & $\Gamma/\Gamma _1$ & $N_{H,z}/\Gamma_2$ & $E_b$ & $\chi ^2/d.o.f.$ & $F_{2-10}$ \\ 
(1)&(2)&(3)&(4)&(5)&(6)\\\hline 
&&&&&\\
pl+GA+zA& $1.24\pm 0.01$ & 1.65$\pm 0.09$& -- & 1033.3/1068& 15.7\\ 
bpl+GA& 0.26$\pm 0.06$& 1.21$\pm 0.01$& 1.07$\pm 0.04$& 1066.3/1067& 15.7\\ 
&&&&&\\ \hline
\end{tabular}
\end{center}
\caption{\scriptsize Best fit parameters for the {\it XMM-Newton} data. Colums as in Tab.1.}
\label{xmm}
\end{table}

\begin{table}[h]
\begin{center}
\begin{tabular}{lccccccccc}\hline
Model & $B$ & $\Gamma _b$ & $K$ & $n_1$ & $n_2$ & $\gamma _{\rm p}$ & $\gamma _{\rm max}$ & $P_j$ &$L_{\rm rad}$\\
& (G) & &(cm$^{-3}$) & & & & & ($10^{48}$ erg/s) & ($10^{47}$ erg/s)\\\hline
2003-{\it C} & 1& 20& $2.45\times 10^4 $&1.5& 3.5& 40& $8\times 10^3$ & 1.9 & 1.2\\
2006-{\it C}, $\gamma $ high & 2.5& 20& $2.45\times 10^4 $&2& 3.5& 100& $1\times 10^4$ & 1.4& 1.2\\
2006-{\it C}, $\gamma $  low & 5.5& 20& $2.45\times 10^4 $&2& 3.5& 20& $1\times 10^4$ & 1.4&0.3\\
2003-{\it A} & 1.55& 13& $10^5 $&1.5& 3.5& 60& $2\times 10^4$ & 3.1 &0.6\\
2006-{\it A}  & 2.5& 13& $1.9\times 10^5 $&2& 3.5& 100& $1\times 10^4$ & 4.4 &1.9\\ \hline
\end{tabular}
\end{center}
\caption{\scriptsize Model parameters used to calculate the SEDs
reported in Fig.(\ref{sedall})-(\ref{sedallabs}). In all the models we
assume a radius $R=2.5\times 10^{16}$ cm, a minimum electron Lorentz
factor $\gamma _{\rm min}=1$, a disk luminosity $L_D=4\times 10^{47}$
erg/s and a BLR radius $R_{BLR}=10^{18}$ cm.}
\label{parameters}
\end{table}

\clearpage

\end{document}